\documentclass[aps,prapplied,reprint,superscriptaddress]{revtex4-2}
\usepackage{amsmath}
\usepackage{amssymb}
\usepackage{graphicx}
\usepackage{dcolumn}
\usepackage{bm}
\usepackage{color}

\begin{document}

\title{Discrete-modulation continuous-variable quantum key distribution with high key rate}

\author{Pu Wang}
\affiliation{State Key Laboratory of Quantum Optics and Quantum Optics Devices, Institute of Opto-Electronics, Shanxi University, Taiyuan 030006, People's Republic of China}
\author{Jianqiang Liu}
\affiliation{State Key Laboratory of Quantum Optics and Quantum Optics Devices, Institute of Opto-Electronics, Shanxi University, Taiyuan 030006, People's Republic of China}
\author{Zhenguo Lu}
\affiliation{State Key Laboratory of Quantum Optics and Quantum Optics Devices, Institute of Opto-Electronics, Shanxi University, Taiyuan 030006, People's Republic of China}
\author{Xuyang Wang}
\affiliation{State Key Laboratory of Quantum Optics and Quantum Optics Devices, Institute of Opto-Electronics, Shanxi University, Taiyuan 030006, People's Republic of China}
\affiliation{Collaborative Innovation Center of Extreme Optics, Shanxi University, Taiyuan 030006, People's Republic of China}
\author{Yongmin Li}
\email{yongmin@sxu.edu.cn}
\affiliation{State Key Laboratory of Quantum Optics and Quantum Optics Devices, Institute of Opto-Electronics, Shanxi University, Taiyuan 030006, People's Republic of China}
\affiliation{Collaborative Innovation Center of Extreme Optics, Shanxi University, Taiyuan 030006, People's Republic of China}

\date{October 1, 2021}

\begin{abstract}
Discrete-modulation continuous-variable quantum key distribution has the potential for large-scale deployment in the secure quantum communication networks due to low implementation complexity and compatibility with the current telecom systems. The security proof for four coherent states phase-shift keying (4-PSK) protocol has recently been established by applying numerical methods. However, the achievable key rate is relatively low compared with the optimal Gaussian modulation scheme. To enhance the key rate of discrete-modulation protocol, we first show that 8-PSK increases the key rate by about 60\% in comparison to 4-PSK, whereas the key rate has no significant improvement from 8-PSK to 12-PSK. We then expand the 12-PSK to two-ring constellation structure with four states in the inner ring and eight states in the outer ring, which significantly improves the key rate to be 2.4 times of that of 4-PSK. The key rate of the two-ring constellation structure can reach 70\% of the key rate achieved by Gaussian modulation in long distance transmissions, making this protocol an attractive alternative for high-rate and low-cost application in secure quantum communication networks.
\end{abstract}

\maketitle
\section{INTRODUCTION}
Quantum key distribution (QKD) \cite{1,2}, as one of the most prominent application of quantum information sciences, allows two distant parties to share a common secret key, where the security is guaranteed by the fundamental laws of quantum physics \cite{3,4}. Continuous-variable (CV) QKD encodes the key information into continuous-spectrum quantum observables, such as the quadrature components of the light field, which has the potential to offer larger key rates at metropolitan distances \cite{5,6,7,8}. CV-QKD can employ the similar components as classical telecom systems and has received extensive attentions and witnessed rapid development both theoretically and experimentally \cite{9,10,11,12,13,14,15,16,17,18,19,20,21,22,23,24,25,
26,27,28,29,30,31,32,33,34,35}.

At present, most of the CV-QKD schemes are based on Gaussian modulation, meaning that Alice displaces the quadratures of the sent states according to a Gaussian distribution, which reaches the channel capacity and achieves a high key rate. However, this type of protocol puts a lot of requirements on the modulation devices and error-correction procedure. Moreover, a perfectly Gaussian modulation cannot be met in realistic applications due to finite range and precision of practice modulators. In practice, the Guassian modulation is approximated by a modulation constellation with a finite number of states and it has shown that at least 8100 states (90 $\times$ 90 size constellation) are needed to satisfactorily simulate a Gaussian distribution \cite{36}. In order to release these stringent restrictions and simplify the protocol, discrete-modulation schemes for CV-QKD are proposed \cite{36,37,38,39,40,41}. Among them, four coherent states phase-shift keying (4-PSK) protocol has attracted the most interest. However, different from Gaussian modulation that can apply the proof method of Gaussian attacks optimality \cite{42,43,44}, the security analysis of discrete modulation CV-QKD is more complex. Previous security proofs for 4-PSK protocol is restricted to Gaussian attacks \cite{45,46,47}, which are not believed to be optimal for discrete modulation schemes, thus the key rate obtained cannot be considered to be secure.

Recently, the security of such protocol in the asymptotic regime is established by applying numerical methods \cite{48,49}, more precisely convex optimization techniques. Without using the Gaussian optimality proof method, the approach in Ref. \cite{49} provides a tighter bound and thus higher key rate than that in Ref. \cite{48}. However, that key rate is still relatively low compared with the optimal Gaussian modulation, about a quarter of the key rate achievable of Gaussian modulation, which makes the 4-PSK protocol less attractive.

In this paper, we focus on the design of the optimized discrete-modulation protocol. To this end, we first extend the 4-PSK protocol to more signal states, eight states (8-PSK) and twelve states (12-PSK), and derive the asymptotic secure key rate by numerical methods considering the realistic trusted noisy detection. The results show that 8-PSK increases the key rate by about 60\%, compared with the original 4-PSK protocol, while the key rate has only small improvement from 8-PSK to 12-PSK. This is because the performance of 8-PSK has tended to be saturated for a single-ring constellation protocol, and further increasing the number of states would not gain more advantages. To enlarge the distribution range of states in phase space and further improve the performance, we propose to use the two-ring constellation scheme, where not only the phase quadrature but the amplitude quadrature is modulated. By optimizing the amplitudes and the probabilities assigned to each states, and the boundary between the two rings, the two-ring constellation with twelve states (four states in the inner ring and eight states in the outer ring) get a superior performance. In comparison with the original 4-PSK protocol, the secret key rate is increased to 2.4 times. This value is about 70\% of the key rate achievable for Gaussian modulation and therefore the performance of the two-ring constellation protocol is close to the Gaussian modulation protocol, but the implementation is simpler, which makes it very attractive for practical applications.

The rest of the paper is organized as follows. In Sec. II, we analyze the performance of discrete-modulation protocol of 8-PSK and 12-PSK with single-ring constellation. In Sec. III, we apply the two-ring constellation structure to 12-PSK, and investigate the dependence of key rate on various parameters to optimize the protocol. The results are compared with Gaussian modulation CV-QKD protocol under realistic trusted noisy detection. Our conclusions are drawn in Sec. IV.

\section{DISCRETE-MODULATION CV-QKD WITH SINGLE-RING CONSTELLATION}
\subsection{The protocol description}
\begin{figure}[htbp]
\centering\includegraphics{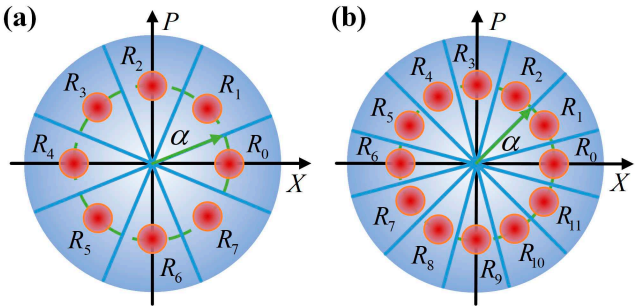}
\caption{\label{Fig_1}Schematic diagram of the single-ring constellation for multiple signal states: (a) 8-PSK, (b) 12-PSK. ${{R}_{j}}$ represents the key mapping region, where Bob maps the measurement outcomes to the corresponding discretized key values.}
\end{figure}

The schematic of single-ring signal constellation for discrete-modulation CV-QKD of 8-PSK and 12-PSK is illustrated in Fig.~\ref{Fig_1}. For 8-PSK, the sender Alice prepares the coherent state $\left| {{\alpha }_{x}} \right\rangle =\left| \alpha {{e}^{{ix\pi }/{4}}} \right\rangle $ with $x\in \left\{ 0,1,2,3,4,5,6,7 \right\}$, each coherent state is chosen with an equal probability ${{p}_{x}}={1}/{8}$. Similarly, for 12-PSK, Alice randomly selects a state from the set $\left\{ \left| {{\alpha }_{x}} \right\rangle =\left| \alpha {{e}^{{ix\pi }/{6}}} \right\rangle ,\text{ }x=0,\cdot \cdot \cdot ,11 \right\}$, where $\alpha $ is a pre-determined amplitude and can be optimized. The prepared states are sent to Bob through an insecure quantum channel to generate secret keys. Bob measures the received states with heterodyne detection and records the measurement outcome $y\in \mathbb{C}$.

After obtaining enough data, Alice and Bob randomly select a small part of the data for parameter estimation. If the parameter estimation passes, they use the remaining data to extract the keys. Bob obtains his raw key string by a key map. Specifically, Bob labels each outcome $y=\left| y \right|{{e}^{i\theta }}$ according to the region ${{R}_{j}}$ as
\begin{equation}
z=\left\{ \begin{array}{l}
   j,\text{if }\theta \in \left[ \frac{\left( 2j-1 \right)\pi }{8},\frac{\left( 2j+1 \right)\pi }{8} \right)\to y\in {{R}_{j}}\text{,  8-PSK} \\
  j,\text{if }\theta \in \left[ \frac{\left( 2j-1 \right)\pi }{12},\frac{\left( 2j+1 \right)\pi }{12} \right)\to y\in {{R}_{j}}\text{,  12-PSK}
\end{array} \right.,
\end{equation}
where $j\in \left\{ 0,\cdot \cdot \cdot ,7 \right\}$ for 8-PSK and $j\in \left\{ 0,\cdot \cdot \cdot ,11 \right\}$ for 12-PSK. The raw key string of Alice consists of the label $x$ of her randomly selected state ${{\alpha }_{x}}$. Finally, Alice and Bob implement error correction and privacy amplification procedures to extract secret keys.
\subsection{Secret key rate}
For convenience of security analysis, we consider the equivalent entanglement-based (EB) scheme. Here, we mainly use the numerical security proof method presented in Ref. \cite{49,50} to obtain the secure key rate. In the EB scheme, Alice initially prepares the bipartite state
\begin{equation}
{{\left| \psi  \right\rangle }_{A{A}'}}=\sum\limits_{x}{\sqrt{{{p}_{x}}}}{{\left| x \right\rangle }_{A}}{{\left| {{\alpha }_{x}} \right\rangle }_{{{A}'}}},
\end{equation}
where $\left| x \right\rangle $ is an orthonormal basis for register $A$. Alice keeps $A$ and sends the register ${A}'$ to Bob. To establish the equivalence between the EB scheme and the original prepare-and-measure scheme, Alice applies a local projective measurement on register $A$, which can be described by a POVM ${{M}^{A}}=\left\{ M_{x}^{A}=\left| x \right\rangle \left\langle  x \right| \right\}$. When obtaining a measurement outcome $x$ with probability ${{p}_{x}}$, the state sent to Bob is effectively collapsed to $\left| {{\alpha }_{x}} \right\rangle $. After the quantum channel transmission, the joint state shared by Alice and Bob is
\begin{equation}
{{\rho }_{AB}}=\left( {{\operatorname{id}}_{A}}\otimes {{\mathcal{E}}_{{A}'\to B}} \right)\left( \left| \psi  \right\rangle {{\left\langle  \psi  \right|}_{A{A}'}} \right),
\end{equation}
where ${{\operatorname{id}}_{A}}$ is the identity channel acting on $A$ and ${{\mathcal{E}}_{{A}'\to B}}$ describes the quantum channel, which is a completely positive and trace-preserving (CPTP) map. Bob uses his POVM ${{G}_{y}}$ on register $B$ to perform realistic trusted noisy detection.

With the reverse reconciliation, the asymptotic secret key rate against collective attacks is expressed as \cite{49}
\begin{equation}
{{K}^{\infty }}=\min \limits_{{{\rho }_{AB}}\in \mathcal{S}}D\left( \left. \mathcal{G}\left( {{\rho }_{AB}} \right) \right\|\mathcal{Z}\left[ \mathcal{G}\left( {{\rho }_{AB}} \right) \right] \right)-{{p}_{\operatorname{pass}}}{{\delta }_{\operatorname{EC}}},
\end{equation}
where $D\left( \left. \rho  \right\|\sigma  \right)=\operatorname{Tr}\left( \rho {{\log }_{2}}\rho  \right)-\operatorname{Tr}\left( \rho {{\log }_{2}}\sigma  \right)$ is the quantum relative entropy; $\mathcal{G}$ describes a completely positive and trace nonincreasing map for postprocessing steps; $\mathcal{Z}$ denotes a pinching quantum channel that reads out the result of key map; $\mathcal{S}$ represents the set of available density operators compatible with experimental observations; ${{p}_{\operatorname{pass}}}$ is the sifting probability of data for key generation; ${{\delta }_{\operatorname{EC}}}$ stands for the leaked information of per signal pulse during the error correction phase and can be computed as follows
\begin{eqnarray}
   {{\delta }_{\operatorname{EC}}}&=&H\left( Z \right)-\beta I\left( X;Z \right) \nonumber \\
 &=&\left( 1-\beta  \right)H\left( Z \right)+\beta H\left( \left. Z \right|X \right),
\end{eqnarray}
where $\beta $ is the reconciliation efficiency. $X$ and $Z$ represent the raw key string of Alice and Bob, respectively.

In order to compute the expected secret key rate, the key point is to find the minimum value of $D\left( \left. \mathcal{G}\left( {{\rho }_{AB}} \right) \right\|\mathcal{Z}\left[ \mathcal{G}\left( {{\rho }_{AB}} \right) \right] \right)$, which is a convex optimization problem matching some constraints and expressed as \cite{50}
\begin{equation}
\begin{aligned}
  & \text{minimize }D\left( \left. \mathcal{G}\left( {{\rho }_{AB}} \right) \right\|\mathcal{Z}\left[ \mathcal{G}\left( {{\rho }_{AB}} \right) \right] \right) \\
 & \text{subject to} \\
 & \quad\quad\quad\quad\operatorname{Tr}\left[ {{\rho }_{AB}}\left( \left| x \right\rangle {{\left\langle  x \right|}_{A}}\otimes {{{\hat{F}}}_{Q}} \right) \right]={{p}_{x}}{{\left\langle {{{\hat{F}}}_{Q}} \right\rangle }_{x}} \\
 & \quad\quad\quad\quad\operatorname{Tr}\left[ {{\rho }_{AB}}\left( \left| x \right\rangle {{\left\langle  x \right|}_{A}}\otimes {{{\hat{F}}}_{P}} \right) \right]={{p}_{x}}{{\left\langle {{{\hat{F}}}_{P}} \right\rangle }_{x}} \\
 & \quad\quad\quad\quad\operatorname{Tr}\left[ {{\rho }_{AB}}\left( \left| x \right\rangle {{\left\langle  x \right|}_{A}}\otimes {{{\hat{S}}}_{Q}} \right) \right]={{p}_{x}}{{\left\langle {{{\hat{S}}}_{Q}} \right\rangle }_{x}} \\
 & \quad\quad\quad\quad\operatorname{Tr}\left[ {{\rho }_{AB}}\left( \left| x \right\rangle {{\left\langle  x \right|}_{A}}\otimes {{{\hat{S}}}_{P}} \right) \right]={{p}_{x}}{{\left\langle {{{\hat{S}}}_{P}} \right\rangle }_{x}} \\
 & \quad\quad\quad\quad\operatorname{Tr}\left[ {{\rho }_{AB}} \right]=1, \\
 & \quad\quad\quad\quad{{\rho }_{AB}}\ge 0 \\
\end{aligned},
\end{equation}
where ${{\hat{F}}_{Q}}$ and ${{\hat{F}}_{P}}$ (${{\hat{S}}_{Q}}$ and ${{\hat{S}}_{P}}$) represent the first-moment (second-moment) observables of quadrature operators $\hat{q}$ and $\hat{p}$, respectively. Also notice that Eve cannot access Alice's system, thus there is an additional constraint
\begin{equation}
{{\operatorname{Tr}}_{B}}\left[ {{\rho }_{AB}} \right]={{\rho }_{A}}=\sum\limits_{i,j=0}^{n}{\sqrt{{{p}_{i}}{{p}_{j}}}}\left\langle  {{\alpha }_{j}} | {{\alpha }_{i}} \right\rangle \left| i \right\rangle {{\left\langle  j \right|}_{A}},
\end{equation}
where $n$ takes 7 or 11 according to 8-PSK or 12-PSK.

The postprocessing map $\mathcal{G}$ and pinching quantum channel $\mathcal{Z}$ can be written as \cite{50}

\begin{equation}
\begin{aligned}
  & \mathcal{G}\left( \sigma  \right)=K\sigma {{K}^{\dagger }}, \\
 & K=\sum\limits_{z=0}^{n}{{{\left| z \right\rangle }_{R}}\otimes {{\operatorname{I} }_{A}}\otimes {{\left( \sqrt{{{R}_{z}}} \right)}_{B}}}, \\
 & \mathcal{Z}\left( \sigma  \right)=\sum\limits_{j=0}^{n}{\left( \left| j \right\rangle {{\left\langle  j \right|}_{R}}\otimes {{\operatorname{I} }_{AB}} \right)\sigma \left( \left| j \right\rangle {{\left\langle  j \right|}_{R}}\otimes {{\operatorname{I}}_{AB}} \right)}. \\
\end{aligned}
\end{equation}
To numerically minimize the objective function, available ${{\rho }_{AB}}$ must be a finite dimension. The dimension of Alice's system is determined by the number of different signal states that she prepares, which is finite. However the receiver Bob's state is in an infinite dimensional Hilbert space. This can be circumvented by applying the photon-number cutoff assumption. When the photon number cutoff parameter ${{N}_{c}}$ is chosen to be large enough, this assumption is reasonable. Here, we set an appropriate parameter value of ${{N}_{c}}$ such that the probability of photon number of the received signal state smaller than ${{N}_{c}}$ is close to 1, and larger ${{N}_{c}}$ has no meaningful improvement for the key rate. Then, the matrix expressions of the first-moment observables ${{\hat{F}}_{Q}}$, ${{\hat{F}}_{P}}$, the second-moment observables ${{\hat{S}}_{Q}}$, ${{\hat{S}}_{P}}$, and the region operators ${{R}_{j}}$ in the photon-number basis can be obtained \cite{50}.

A tight lower bound on the key rate can be achieved by using a two-step procedure developed in Ref. \cite{51}. At the first step, the Frank-Wolfe algorithm is used to approximately minimize the convex function, and hence obtaining an upper bound on the key rate. At the second step, converting this upper bound to a reliable lower bound by taking the numerical imprecision into account. By using the two-step procedure and defining $f\left( \rho  \right)=D\left( \left. \mathcal{G}\left( \rho  \right) \right\|\mathcal{Z}\left[ \mathcal{G}\left( \rho  \right) \right] \right)$, we have
\begin{equation}
\min\limits_{\rho \in \mathcal{S}}f\left( \rho  \right)\ge {{f}_{\epsilon }}\left( {{\rho }_{1}} \right)-\operatorname{Tr}\left( \rho _{1}^{T}\nabla {{f}_{\epsilon }}\left( {{\rho }_{1}} \right) \right)+f_{d}^{\max }-{{\zeta }_{\epsilon }},
\end{equation}
where ${{\rho }_{1}}$ is the suboptimal state obtained in the first step. $\epsilon $ is the perturbation parameter and ${{f}_{\epsilon }}\left( {{\rho }_{1}} \right):=D\left( \left. {{\mathcal{G}}_{\epsilon }}\left( {{\rho }_{1}} \right) \right\|\mathcal{Z}\left[ {{\mathcal{G}}_{\epsilon }}\left( {{\rho }_{1}} \right) \right] \right)$. ${{\zeta }_{\epsilon }}=2\epsilon \left( {d}'-1 \right){{\log }_{2}}\frac{{{d}'}}{\epsilon \left( {d}'-1 \right)}$, where ${d}'$ is the dimension of $\mathcal{G}\left( \rho  \right)$.
The gradient $\nabla {{f}_{\epsilon }}\left( {{\rho }_{1}} \right)$ is given by
\begin{equation}
{{\left[ \nabla {{f}_{\epsilon }}\left( {{\rho }_{1}} \right) \right]}^{T}}=\mathcal{G}_{\epsilon }^{\dagger }\left( {{\log }_{2}}{{\mathcal{G}}_{\epsilon }}\left( {{\rho }_{1}} \right) \right)-\mathcal{G}_{\epsilon }^{\dagger }\left( {{\log }_{2}}\mathcal{Z}\left[ {{\mathcal{G}}_{\epsilon }}\left( {{\rho }_{1}} \right) \right] \right).
\end{equation}
$f_{d}^{\max }$ is the dual function and written as
\begin{equation}
\begin{aligned}
  & f_{d}^{\max }=\underset{\left( \vec{v},\vec{s} \right)}{\mathop{\max }}\,\left( \vec{\tilde{\gamma }}\cdot \vec{v}-{\epsilon }'\sum\limits_{i=1}^{nc}{{{s}_{i}}} \right) \\
 & \text{subject to} \\
 &\quad\quad\quad\quad -\vec{s}\le \vec{v}\le \vec{s} \\
 &\quad\quad\quad\quad \sum\limits_{i=1}^{nc}{{{v}_{i}}\tilde{\Gamma }_{i}^{T}}\le \nabla {{f}_{\epsilon }}\left( {{\rho }_{1}} \right) \\
 &\quad\quad\quad\quad \left( \vec{v},\vec{s} \right)\in \left( {{\mathbb{R}}^{nc}},{{\mathbb{R}}^{nc}} \right) \\
\end{aligned},
\end{equation}
where ${{\Gamma }_{i}}$ and ${{\gamma }_{i}}$ refer to both sides of the equality constraint, respectively, with the form $\operatorname{Tr}\left( {{\Gamma }_{i}}\rho  \right)={{\gamma }_{i}}$. $nc$ represents the number of all constraints. ${\epsilon }'$ is the security parameter related to the numerical imprecision.

We are now in a position to assess the performance of the protocol based on the above approach.
\subsection{Performance analysis}
Considering a typical phase-insensitive Gaussian channel in the context of the optical fiber communication. The simulated statistics are given by \cite{50}
\begin{equation}
\begin{aligned}
  & {{\left\langle {{{\hat{F}}}_{Q}} \right\rangle }_{x}}=\sqrt{2\eta T}\operatorname{Re}\left( {{\alpha }_{x}} \right), \\
 & {{\left\langle {{{\hat{F}}}_{P}} \right\rangle }_{x}}=\sqrt{2\eta T}\operatorname{Im}\left( {{\alpha }_{x}} \right), \\
 & {{\left\langle {{{\hat{S}}}_{Q}} \right\rangle }_{x}}=2\eta T\operatorname{Re}{{\left( {{\alpha }_{x}} \right)}^{2}}+1+\frac{1}{2}\eta T\xi +{{v}_{el}}, \\
 & {{\left\langle {{{\hat{S}}}_{P}} \right\rangle }_{x}}=2\eta T\operatorname{Im}{{\left( {{\alpha }_{x}} \right)}^{2}}+1+\frac{1}{2}\eta T\xi +{{v}_{el}}, \\
\end{aligned}
\end{equation}
where $T$ and $\xi $ represent the transmittance and excess noise of the channel, respectively, and $\eta $ and ${{v}_{el}}$ denote the detection efficiency and electronic noise of the detector.

The probability density function for the result $y$ of a heterodyne measurement conditioned on Alice's choice $x$ is
\begin{equation}
P\left( \left. y \right|x \right)=\frac{1}{\pi \left( 1+\frac{1}{2}\eta T\xi +{{v}_{el}} \right)}\exp \left[ -\frac{{{\left| y-\sqrt{\eta T}{{\alpha }_{x}} \right|}^{2}}}{1+\frac{1}{2}\eta T\xi +{{v}_{el}}} \right].
\end{equation}
Then, according to the key mapping area, the probability that Bob obtains the discretized key value $Z=z$ conditioned on Alice's choice $x$ is
\begin{equation}
\tilde{P}\left( \left. z \right|x \right)=\left\{ \begin{array}{l}
   \int_{0}^{\infty }{rdr}\int_{{\left( 2z-1 \right)\pi }/{8}}^{{\left( 2z+1 \right)\pi }/{8}}{P\left( \left. r{{e}^{i\theta }} \right|x \right)}d\theta \text{  (8-PSK)} \\
  \int_{0}^{\infty }{rdr}\int_{{\left( 2z-1 \right)\pi }/{12}}^{{\left( 2z+1 \right)\pi }/{12}}{P\left( \left. r{{e}^{i\theta }} \right|x \right)}d\theta \text{ (12-PSK)}
\end{array} \right..
\end{equation}

\begin{figure}[htbp]
\centering
\begin{minipage}{8.6cm}
\centering
\includegraphics{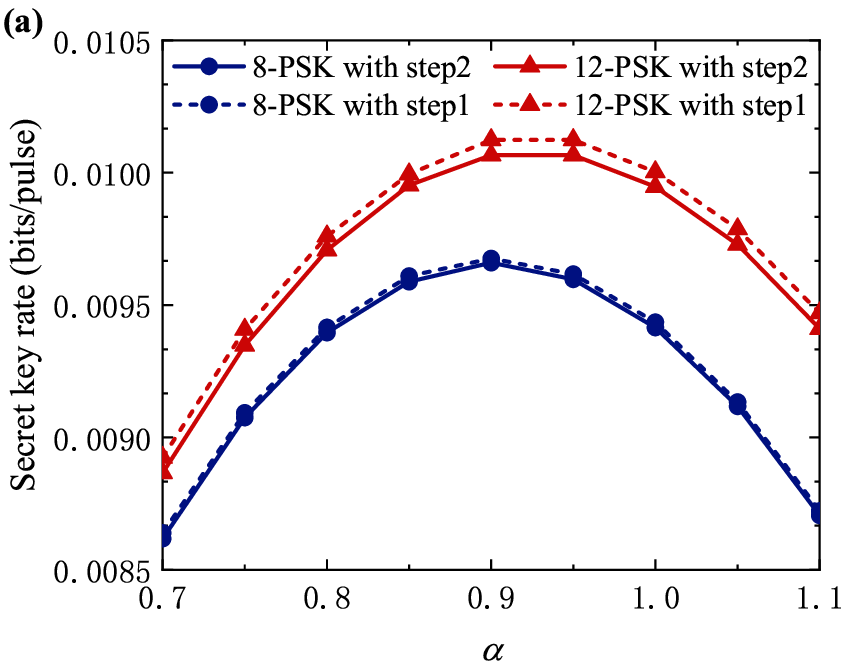}
\end{minipage}
\;
\begin{minipage}{8.6cm}
\centering
\includegraphics{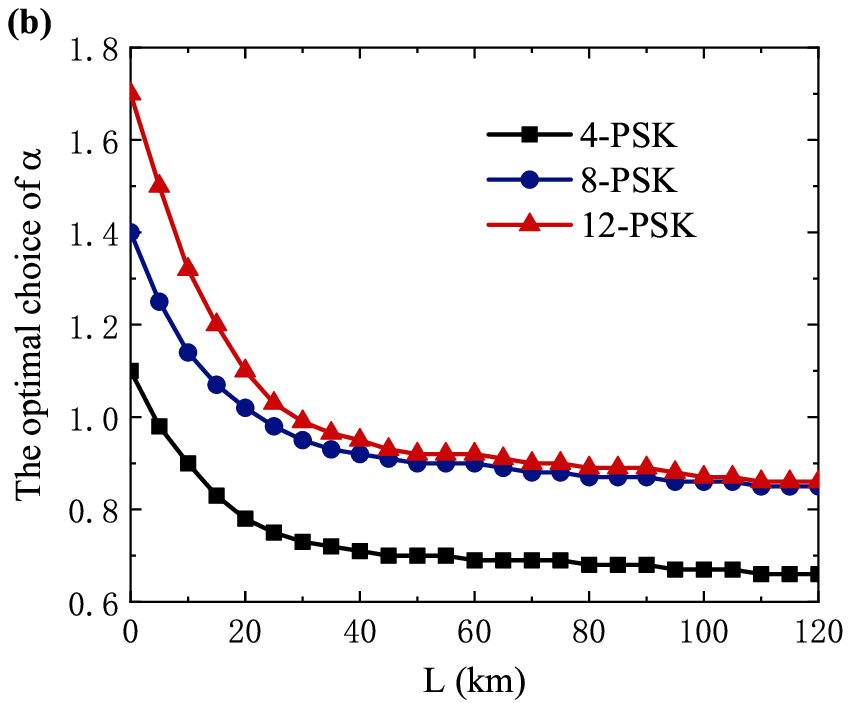}
\end{minipage}
\caption{\label{Fig_2}(a) Secret key rates as a function of the coherent state amplitude $\alpha $ for 8-PSK and 12-PSK protocols at the transmission distance of $L=50$ km. (b) The optimal choice of $\alpha $ at different transmission distances for 4-PSK, 8-PSK and 12-PSK protocols.}
\end{figure}

To extract the optimal secret key rate and fairly compare the performance of different protocols, we first investigate the dependence of the key rate on the coherent state amplitude $\alpha $. In Fig.~\ref{Fig_2}(a), we plot the key rate versus the choice of $\alpha $ for 8-PSK and 12-PSK protocols, at a transmission distance of 50 km. Here, we consider the realistic parameters: reconciliation efficiency $\beta =0.95$, excess noise $\xi =0.01$, detection efficiency $\eta =0.552$, and electronic noise ${{v}_{el}}=0.015$. The dashed line represents the result of step 1 of our algorithm, which gives an upper bound on the key rate, and solid line is the result after step 2, which provides an achievable lower secure bound on the key rate. The gap between step 1 and step 2 is small, hence the obtained bound is tight. We carry out a coarse-grained search for $\alpha $ in the interval [0.7, 1.1] with a step size of 0.05. It is clear that there is an optimal value of $\alpha $ to maximize the key rate. At 50 km, the optimal $\alpha $ for 8-PSK and 12-PSK is 0.9 and 0.92, respectively.  In Fig.~\ref{Fig_2}(b), we give the optimal values of $\alpha $ at different transmission distances and the results of 4-PSK protocol are also shown for comparison. Compared with the 4-PSK protocol, 8-PSK allows larger optimal value of $\alpha $, thus providing higher signal-to-noise ratio (SNR). The difference between the optimal $\alpha $ for the 8-PSK and 12-PSK protocols gradually decreases with the increase of transmission distance and is almost indistinguishable over long distances.

\begin{figure}[htbp]
\centering\includegraphics{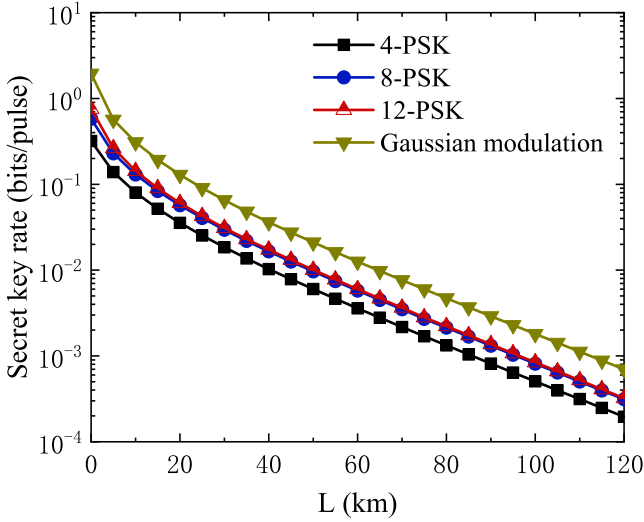}
\caption{\label{Fig_3}Achievable secret key rate versus the transmission distance for different protocols. The modulation amplitude and variance have been optimized. Other parameters are set to $\beta =0.95$, $\xi =0.01$, $\eta =0.552$, and ${{v}_{el}}=0.015$.}
\end{figure}

In Fig.~\ref{Fig_3}, we present the achievable key rates at different transmission distances for different discrete-modulation protocols and Gaussian modulation protocol. The key rate has been optimized by adopting the optimal modulation amplitude (Fig.~\ref{Fig_2}(b)) and variance. In addition, we have applied the postselection technology to the 4-PSK protocol to obtain the optimal key rate \cite{50}. We can see that the 8-PSK protocol improves the key rate by about 60\% over the 4-PSK protocol. Within 10 km, the key rate can be further improved by 10\% - 30\% if 12-PSK is employed. However, beyond 10 km, the key rate can only be increased by about 4\%, which means that the performance of 8-PSK modulation is close to saturation and adding more signal states will not make much difference. Compared with the Gaussian-modulation protocol, the achievable key rate of discrete-modulation is still relatively low. Taking 50 km as an example, the key rates per pulse of the 4-PSK, 8-PSK, 12-PSK and Gaussian modulation protocols are 0.00602, 0.00966, 0.01008, and 0.02103 bits/pulse, respectively. The 4-PSK and 8-PSK protocols can reach approximately 28\% and 45\% of the key rate of Gaussian modulation. The data postselection is also an interesting strategy to further improve the key rate by 8\% for 8-PSK and 12-PSK protocols (see Appendix).
\section{DISCRETE-MODULATION CV-QKD WITH TWO-RING CONSTELLATION }
\begin{figure}[htbp]
\centering\includegraphics[width=7.6cm]{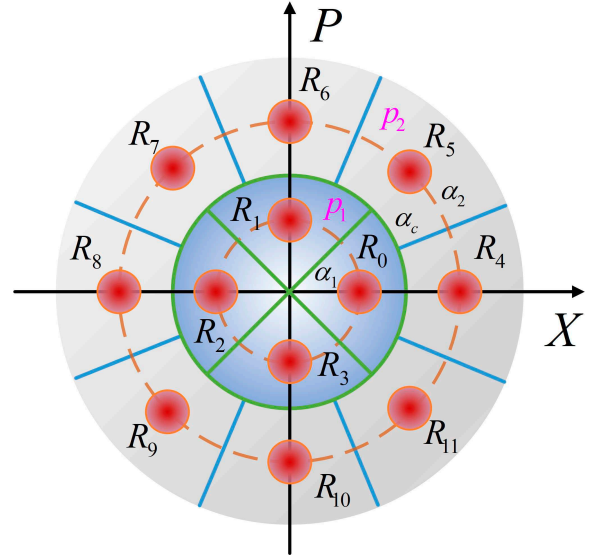}
\caption{\label{Fig_4}Schematic diagram of the proposed two-ring constellation structure with twelve signal states.}
\end{figure}

To further improve the performance of the discrete-modulation protocol, we extend the 12-PSK to two-ring constellation structure. As shown in Fig.~\ref{Fig_4}, different from the single-ring PSK protocol, the states are prepared with two different amplitudes ${{\alpha }_{1}}$ and ${{\alpha }_{2}}$ in the designed two-ring constellation. The four states in the inner ring are expressed as ${{\left\{ \left| {{\alpha }_{x}} \right\rangle =\left| {{\alpha }_{1}}{{e}^{{ix\pi }/{2}}} \right\rangle  \right\}}_{x=0,\cdot \cdot \cdot ,3}}$, where each of states is chosen according to an equal probability ${{p}_{1}}$. The eight states in the outer ring take the form of ${{\left\{ \left| {{\alpha }_{x}} \right\rangle =\left| {{\alpha }_{2}}{{e}^{{i\left( x-4 \right)\pi }/{4}}} \right\rangle  \right\}}_{x=4,\cdot \cdot \cdot ,11}}$ and each of which is chosen with an equal probability ${{p}_{2}}$. ${{p}_{1}}+2{{p}_{2}}={1}/{4}$. Alice transmits the randomly selected state ${{\alpha }_{x}}$ to Bob and records the sequence of the state she sent. Upon receiving the state, Bob performs a heterodyne detection and obtains the measurement outcome $y$. According to the region ${{R}_{j}}$, Bob maps his outcome $y=\left| y \right|{{e}^{i\theta }}$ to the discretized raw keys as follows
\begin{eqnarray}
&&z= \nonumber\\
&&\left\{ \begin{array}{l}
   j,\text{ if }\theta \in \left[ \frac{\left( 2j-1 \right)\pi }{4},\frac{\left( 2j+1 \right)\pi }{4} \right)\text{,}\left| y \right|\in \left[ 0,{{\alpha }_{c}} \right)\text{   }{{\left\{ {{R}_{j}} \right\}}_{0\le j\le 3}} \\
  j,\text{ if }\theta \in \left[ \frac{\left( 2j-9 \right)\pi }{8},\frac{\left( 2j-7 \right)\pi }{8} \right),\left| y \right|\in \left[ {{\alpha }_{c}},\infty  \right)\text{ }{{\left\{ {{R}_{j}} \right\}}_{4\le j\le 11}}
\end{array} \right., \nonumber\\
\end{eqnarray}
where ${{\alpha }_{c}}$ is the amplitude corresponding to the boundary between the inner and outer regions.

\begin{figure}[htbp]
\centering\includegraphics{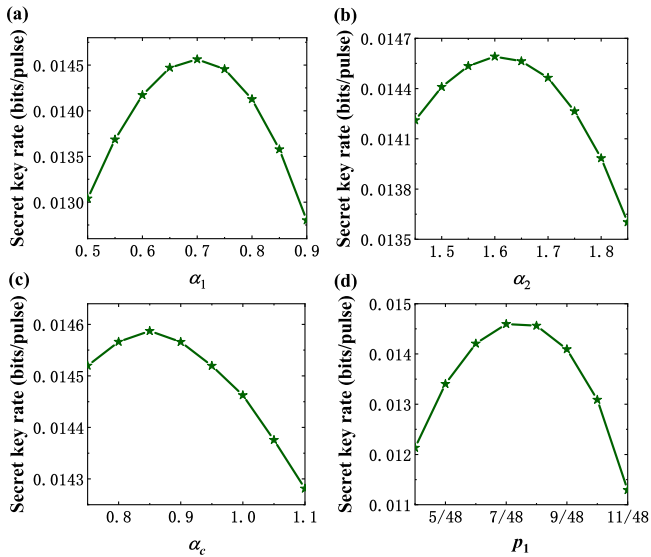}
\caption{\label{Fig_5}Secure key rate versus the choices of ${{\alpha }_{1}}$, ${{\alpha }_{2}}$, ${{\alpha }_{c}}$, and ${{p}_{1}}$. The simulation parameters are $\beta =0.95$, $\xi =0.01$, $\eta =0.552$, ${{v}_{el}}=0.015$, and $L=50\text{ }\operatorname{km}$. (a) The optimal ${{\alpha }_{1}}$ is obtained via a coarse-grained search in the interval [0.5, 0.9] with a step size 0.05. The other parameters are set to ${{\alpha }_{2}}=1.6$, ${{\alpha }_{c}}=0.85$, and ${{p}_{1}}={7}/{48}$. (b) The value of ${{\alpha }_{2}}$ is changed with a step size 0.05 in the interval [1.45, 1.85]. The other parameters are set to ${{\alpha }_{1}}=0.7$, ${{\alpha }_{c}}=0.85$, ${{p}_{1}}={7}/{48}$.  (c) The value of ${{\alpha }_{c}}$ is changed with a step size 0.05 in the interval [0.75, 1.1]. The other parameters are set to ${{\alpha }_{1}}=0.7$, ${{\alpha }_{2}}=1.6$, ${{p}_{1}}={7}/{48}$.  (d) The value of ${{p}_{1}}$ is searched in the interval [4/48, 11/48] with a step size 1/48. The other parameters are set to ${{\alpha }_{1}}=0.7$, ${{\alpha }_{2}}=1.6$, and ${{\alpha }_{c}}=0.85$. }
\end{figure}

The key rate can be optimized by finding the optimal parameter values of ${{\alpha }_{1}}$, ${{\alpha }_{2}}$, ${{\alpha }_{c}}$, and ${{p}_{1}}$ (${{p}_{1}}+2{{p}_{2}}={1}/{4}$). The dependence of key rate on these parameters are shown in Fig.~\ref{Fig_5}. For the transmission distance of 50 km, the optimal choices of the parameters are ${{\alpha }_{1}}=0.7$, ${{\alpha }_{2}}=1.6$, ${{\alpha }_{c}}=0.85$, ${{p}_{1}}={7}/{48}$ and ${{p}_{2}}={5}/{96}$. In this case, we achieve a key rate of 0.01459 bits/pulse, which is 50\% higher than that of 8-PSK protocol and 140\% higher than that of 4-PSK protocol. By varying the constellation geometry from the single-ring PSK structure to the two-ring constellation structure, we effectively overcomes the limitation of states distribution in phase space, and significantly improves the performance of the 4-PSK discrete-modulation protocol.

\begin{figure}[htbp]
\centering\includegraphics{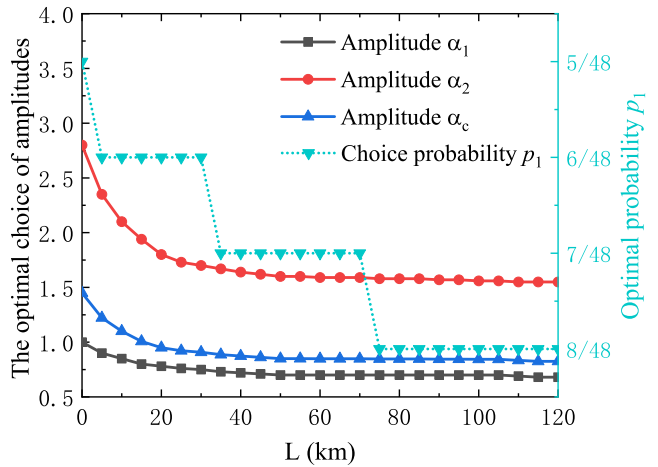}
\caption{\label{Fig_6}The optimal choice of parameters ${{\alpha }_{1}}$, ${{\alpha }_{2}}$, ${{\alpha }_{c}}$ and ${{p}_{1}}$ at different transmission distances for two-ring constellation protocol with twelve signal states.}
\end{figure}

In Fig.~\ref{Fig_6}, we give the optimal choice of parameters ${{\alpha }_{1}}$, ${{\alpha }_{2}}$, ${{\alpha }_{c}}$ and ${{p}_{1}}$ at different transmission distances. The cutoff value of photon number ${{N}_{c}}$ is selected from the interval [14, 22] according to the modulation amplitudes. The optimal ${{p}_{1}}$ is obtained via a coarse-grained search in the interval [1/48, 11/48] with a step size 1/48. We can see that the optimal ${{\alpha }_{c}}$ is closer to ${{\alpha }_{1}}$ and ${{p}_{1}}$ is greater than ${{p}_{2}}$, which result in an approximate Gaussian distribution in phase space. Next, we employ the optimal parameters values to obtain the optimal key rate and analyze the performance of the protocol.

\begin{figure}[htbp]
\centering\includegraphics{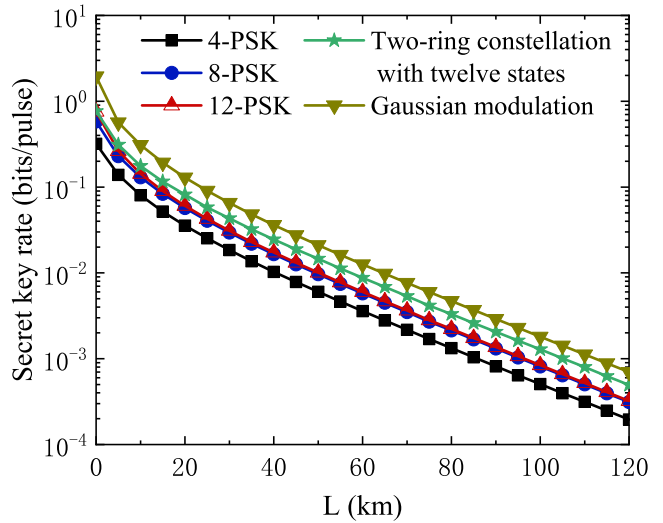}
\caption{\label{Fig_7}Achievable secret key rate versus the transmission distance for different protocols. Other parameters are set to $\beta =0.95$, $\xi =0.01$, $\eta =0.552$, and ${{v}_{el}}=0.015$.}
\end{figure}

\begin{figure}[htbp]
\centering\includegraphics{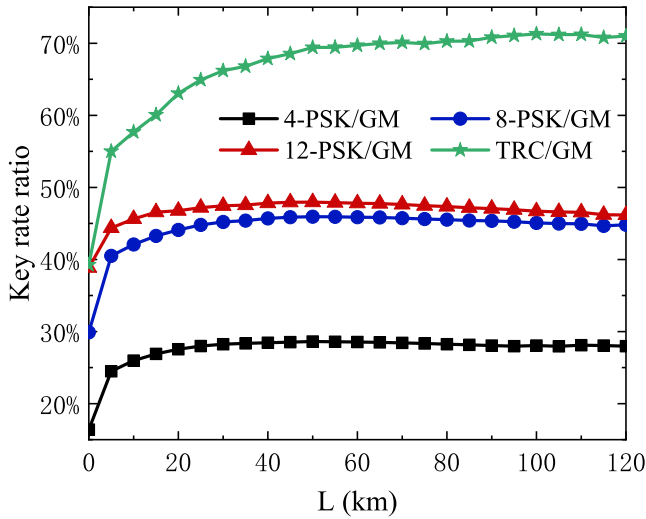}
\caption{\label{Fig_8}Ratio of the key rate of various discrete-modulation protocols to Gaussian modulation protocol. TRC represents two-ring constellation protocol with twelve states and GM represents the Gaussian modulation protocol.}
\end{figure}

In Fig.~\ref{Fig_7}, we compare the key rate of the two-ring constellation protocol with the previous PSK protocols and Gaussian modulation protocol. We observe that the key rate of the two-ring constellation with twelve states is higher than that of PSK protocols, and the gap of the key rate between two-ring constellation protocol and Gaussian modulation protocol decreases gradually with the increase of transmission distance. As shown in Fig.~\ref{Fig_8}, for long distance transmission distance above 50 km, the two-ring constellation with twelve states can reach above 70\% of key rate of Gaussian modulation protocol. Notice that two-ring constellation with twelve states require only 1-bit discretization for the amplitude of the light field and 3-bit discretization for the phase of the light field. Therefore, we proof that the concise two-ring constellation with twelve states can exhibit superior performance, which is very close to that of Gaussian modulation protocol.
\section{CONCLUSIONS}
The optimized constellation design schemes were proposed to significantly improve the performance of the discrete-modulation CV-QKD protocol. The asymptotic secure key rate in the trusted detector noise scenario is obtained by applying the advanced numerical methods. For the sake of fair comparison, we optimized the coherent state amplitude at each transmission distance. Our numerical results showed that 8-PSK can increase the key rate by about 60\% over conventional 4-PSK protocol, however adding the 8-PSK to 12-PSK does not significantly improve the key rate. Interestingly, by allocating twelve coherent states to a two-ring constellation that enlarges the distribution range of coherent states in phase space, the key rate with 140\% higher than that of 4-PSK protocol was achieved. This key rate reaches about 70\% of that of the Gaussian modulation protocol for transmission distance above 50 km. Our results confirm that the discrete-modulation protocol with ten or so appropriate constellation and optimal modulation parameters can approximate the key rate of Gaussian modulation protocol. We believe that the proposed discrete-modulation protocol can find promising applications in high-rate and low-cost secure quantum communication networks.

The future works include the optimization of the key rate of the discrete-modulation protocol in the short distance transmission by extending current state constellation or other new constellation design schemes \cite{52}, and finding better data postselection strategies \cite{53,54}. Extending our security analysis to include the impact of finite-size is also an important future task \cite{55,56,57}.
\begin{acknowledgments}
This work is supported by the National Natural Science Foundation of China (NSFC) (Grants No.11774209); the Aeronautical Science Foundation of China (20200020115001); Key Research and Development Program of Guangdong Province (2020B0303040002); Shanxi 1331KSC.
\end{acknowledgments}
\appendix
\section{POSTSELECTION}
\begin{figure}[htbp]
\centering\includegraphics[width=7.6cm]{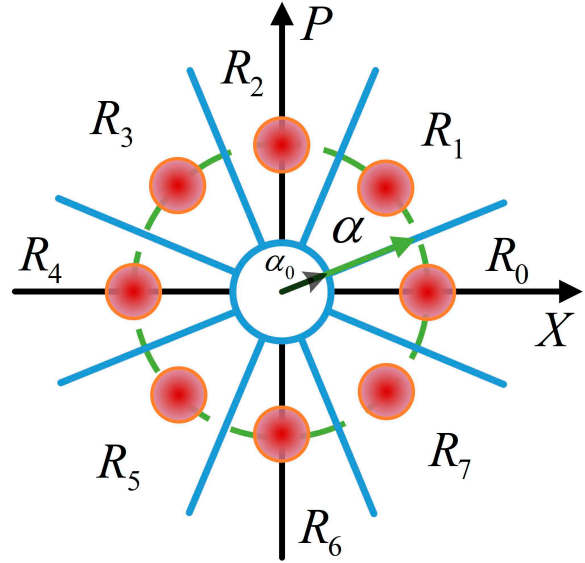}
\caption{\label{Fig_9}The key mapping for 8-PSK protocol when postselection is considered.}
\end{figure}

In this Appendix, we apply the postselection technology to our proposed protocol. Postselection is useful to improve the key rate of the four state protocol \cite{50}. It also reduces the amount of data for post-processing. In Fig.~\ref{Fig_9}, taking the 8-PSK as an example, we illustrate the concept of postselection. There is a cut-off area at the center of the phase space, which is defined by a circle with radius ${{\alpha }_{0}}$. The data inside the circle is discarded and the data outside the circle is retained. Hence, the key mapping results of Bob become
\begin{equation}
z=\left\{ \begin{array}{l}
   j,\text{   if }\theta \in \left[ \frac{\left( 2j-1 \right)\pi }{8},\frac{\left( 2j+1 \right)\pi }{8} \right)\text{ and }\left| y \right|\ge {{\alpha }_{0}}\text{ } \\
  \bot ,\text{   otherwise}
\end{array} \right.,
\end{equation}
where ${{\alpha }_{0}}$ is the postselection parameter, and ${{\alpha }_{0}}=0$ corresponds to the protocol without postselection.

The sifting factor is defined as ${{p}_{\operatorname{pass}}}=\sum\limits_{x,z}{\tilde{P}\left( x \right)\tilde{P}\left( \left. z \right|x \right)}$, where
\begin{equation}
\tilde{P}\left( \left. z \right|x \right)=\int_{{{\alpha }_{0}}}^{\infty }{rdr}\int_{{\left( 2z-1 \right)\pi }/{8}}^{{\left( 2z+1 \right)\pi }/{8}}{P\left( \left. r{{e}^{i\theta }} \right|x \right)}d\theta,
 \label{eq-1}
\end{equation}
$\tilde{P}\left( x \right)$ is the probability that Alice chooses to send the state ${{\alpha }_{x}}$.

\begin{figure}[htbp]
\centering\includegraphics{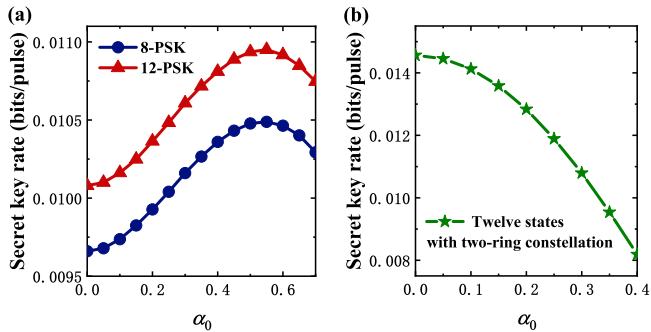}
\caption{\label{Fig_10}(a) Secure key rate versus postselection parameter for the 8-PSK and 12-PSK protocols at the transmission distance of 50 km. The modulation amplitude is set to $\alpha =0.9$ for 8-PSK protocol and $\alpha =0.92$ for 12-PSK protocol. (b) Secure key rate versus postselection parameter for the two-ring constellation protocol at 50km, where the parameters value are set to ${{\alpha }_{1}}=0.7$, ${{\alpha }_{2}}=1.6$, ${{\alpha }_{c}}=0.85$ and ${{p}_{1}}={7}/{48}$. The optimal postselection parameter is obtained via a coarse-grained search for ${{\alpha }_{0}}$ in the interval [0, 0.7] for (a) and [0, 0.4] for (b) with a step size 0.05. Other parameters for simulations are set to $\beta =0.95$, $\xi =0.01$, $\eta =0.552$, and ${{v}_{el}}=0.015$.}
\end{figure}

In Fig.~\ref{Fig_10}(a), we plot the key rate versus the postselection parameter ${{\alpha }_{0}}$ for the 8-PSK and 12-PSK protocols at the distance $L=50$ km. We observe that there exists an optimal postselection parameter to maximize the key rate. The optimal postselection parameter value for 8-PSK and 12-PSK protocols is ${{\alpha }_{0}}=0.55$. Compared with the original protocol without postselection, the key rate is increased by 8\% for both 8-PSK and 12-PSK protocols. By inserting ${{\alpha }_{0}}=0.55$ into Eq.~(\ref{eq-1}), we can obtain the value of the sifting factor ${{p}_{\operatorname{pass}}}=0.75$, which means that the amount of data used for post-processing is reduced by 25\%.

With the similar procedure, we search for the optimal postselection parameter for the two-ring constellation protocol with twelve states, as shown in Fig.~\ref{Fig_10}(b). The result suggests that the optimal value is ${{\alpha }_{0}}=0$. Therefore, we do not need to postselect the data for the two-ring constellation scheme. It should be noted that there may be other postselection strategies that can improve the performance of the protocol, which is left for future work.

\end{document}